# First Principles Study on the Formation of Yttrium Nitride in Cubic and Hexagonal Phases

G. Soto, M.G. Moreno-Armenta and A. Reyes-Serrato

*Centro de Ciencias de la Materia Condensada, Universidad Nacional Autónoma de México, Apartado Postal 356, Ensenada, Baja California. CP 22800, México.*

**Abstract**

We have studied the formation of yttrium nitride in which the Y-atoms were arranged in two common close-packed stacking: The AB-stacking (*hcp*-symmetry) is the ground state of metallic yttrium; while the ABC-stacking (*fcc*-symmetry) is the ground state of yttrium nitride. Given the different symmetries between YN and Y, there must be a phase transition (*hcp* → *fcc*) as N is incorporated in the Y-lattice. By means of first principles calculations we have made a systematical investigation where N was gradually incorporated in octahedral interstices in AB and ABC stacking of Y. We report the heat of formation, bulk modulus, lattice parameter and electronic structure of the resultant nitrides. We found that both metal arrangements are physically achievable. Furthermore, for low nitrogen incorporation the two phases may coexist. However for high nitrogen concentration the cubic phases are favored by a 30 kJ mol$^{-1}$ margin. These results are important since *fcc*-YN is semiconducting and could be utilized as active layer in electronic devices.

**Introduction**

The Transition (T) Metal (M) Nitrides (N) are valuable for several technological applications. Because of their great strength and durability their long established use is as protective coatings [1]. However, they also have interesting optical, electronic, catalytic and magnetic properties [1, 2]. In the electronic industry they are important as electric contact, diffusion barriers, buffer layer, among other uses. In the last years there has been an increased awareness in scandium, yttrium and lanthanum nitrides due to their semiconducting properties [3-9]. In contrast with traditional semiconductors the group IIIB TMN is interstitial alloys; thereafter they could have a wide array of chemical compositions [10-13]. By means of nitrogen and/or metal vacancies their electronic structures may be 'tuned' to specific applications. In future these compounds possibly will be used as the electronically active layers in diodes and transistors.

The structures of group IIIB metal and their nitrides are very simple: face centered cubic (*fcc*), or hexagonal closed packed (*hcp*) [2]. The N atoms are located in interstitial sites of the M-lattice in octahedral geometry [12]. The contrast between *hcp* and *fcc* arrangements of M is the presence of octahedral sites sharing common faces in the first structure, while they share common edges in the second. A TMN in an ideal closed packed arrangement the distance between two N-atoms sharing a common face is the shortest possible: $0.82 \times d$(M-M), whereas for two octahedral sites sharing a common edge the distance is $d$(M-M). Therefore, the interaction between two N-atoms which occupy simultaneously two octahedral sites in an *hcp* arrangement should be higher than in the *fcc* arrangement [13]. The ionic character of the N-atoms explains to some extent the preference for the cubic symmetry in quite a lot TMN. However, is unclear the point in which the N-N interactions supersede the tendency of the parent metal for the *hcp* arrangement [14]. Yttrium nitride is a good study case. It has been reported with a NaCl-type structure, while their parent metal has its ground state in hexagonal symmetry. Therefore, the hexagonal-cubic phase transition is expected in the Y-N systems.

The aim of this work is to study the electronic structure of yttrium nitrides in both, *hcp* and *fcc* phases. At the same time, the calculated heat of formation energies allow to discern the point of phase transition as N is incorporated in edge- and face-sharing octahedral sites of Y. As it will be shown, *fcc*-YN is semiconductor, while our calculations show metallic character for *hcp*-YN.

**Calculation method**

*a) Cell transformations.* The cubic YN structure can be envisaged as two interpenetrating *fcc* sublattices were one sublattice is formed by Y-atoms and the other by N-atoms. This structure is rendered in the *Fm3m* (225) spatial group, with one sublattice in the 4*a* (0, 0, 0) Wyckoff position and the other in the 4*b* (½, ½, ½). The N-atoms fill the octahedral interstices of the Y-lattice, but also the Y-atoms lie in the octahedral interstices of the N-lattice. The next-neighbor octahedron of the same type shares a common edge with it. The stacking of both atoms is ABC. In the other hand, yttrium has an AB stacking. It is described in the *P63/mmc* (194) spatial group using the 2*c* (1/3, 2/3, 1/4) Wyckoff position. In this structure the octahedral interstices lay in the 2*a* (0, 0, 0) Wyckoff position. The octahedral interstices of *hcp*-Y share a common face with their next-neighbor octahedron.

In order to make calculations where the ratio of N to Y could be variable, is essential to reduce the symmetry of both *Fm3m* and *P63/mmc*, groups. In our case we did a series of customized cells in which the cell volumes has been doubled. The symmetry of *Fm3m* was reduced to the *R-3m* (166), Fig. 1(a), and the *P63/mmc* was decreased to *P-3m1* (164), Fig. 1(b). In Table I we summarize the steps to build cells of lower symmetries. These procedures were made using the *subgroup* and *supergroup* utilities in the *PowderCell* program [15]. In the first step we identify the base cell and the Wyckoff position in which the two sublattices are located. In the second step we made larger cells in the same spatial group as the base cell. In the third step we reduce the volume and the symmetry of the expanded cell, bearing in mind that we need the minimum volume and maximum symmetry to obtain good computational times; but simultaneously, keeping the number of unlike points practical for calculations. In Table II we display the stoichiometric points that are useable in the $0 \leq x \leq 1$ interval for the selected groups. The interpenetrating sublattices in the *R-3m* group are identified by 8 Wyckoff position: 6*h* and 2*c* for one sublattice; and 1*a*, 1*b*, 3*d* and 3*e* for the second sublattice. In these positions is possible to set 8 Y atoms and 8 N atoms. For the *P-3m1* group the metallic sites can be situated in 2*d* and 6*i* Wyckoff positions, while the octahedral interstices are in 1*a*, 1*b*, 3*e* and 3*f* positions. Tables I and II resume this information.

*b) Structural and electronic calculations.* For the structural refinement and electronic calculations we have employed the WIEN2k code [16]. This is within the framework of density-functional theory. We have employed the full potential linearized augmented plane wave (FP-LAPW) method as it is implemented in the code. The exchange and correlations effects were treated using the generalized gradient approximation (GGA) of Perdew *et al* [17]. The calculations for cubic and hexagonal phases have been done using an $R_{mt}K_{max}$=7 and $G_{max}$=12. The maximum *l* for the expansion of the wave functions in spherical harmonics inside of nonoverlapping atomic spheres was $l_{max}$=10. We take an energy cut off of -6.0 Ry to separate the core from the valence states. The number of *k*-points used (in the irreducible wedge of the Brillouin zone) was 110 for cubic and 100 for hexagonal phases.

**Results**

Our study is founded in a methodical variation of $x_i$, where $x_i$ was defined as the ratio between N and Y in the *i*-cell; see Table III. As a first step, we calculated the equilibrium structural parameters of Y in hexagonal and cubic symmetries for the extended cells (8 Y-atoms per cell) by means of fitting the Murnaghan equation of state to the energy *vs*. deformation curves; this is the $x_0$-point. The next step was to introduce one nitrogen atom into the metallic lattice. This structure can be considered as a nitrogen-perturbed yttrium lattice, $x_1$-points. Accordingly, the initial guess for cell parameter is the unperturbed, $x_0$, cell parameter. The next point, $x_2$, is generated by introducing two nitrogen atoms. The idea is to recursively obtain all the $x_i$-points of Table II as an approximation to previous $x_{i-1}$ points. The formation energy and structural parameters are summarized in Table III. The $E_i^{tot}$ data for cubic YN and hexagonal Y agree reasonably with calculations [6] and experimental data [18-19]. The behavior of the bulk modulus and cell volume ($V/V_0$) for Y-N series is illustrated in Fig. 2. The bulk modulus shows a steady increase as a function of $x$ as far as $x = 1$. At the same time, the cell volume decreases as $x$ increases.

To compare the stabilities of the various structures we used the formation energy, according to the following definition [20]:

$$E^f = \frac{E_i^{tot} - nE_{Y\_hex}^{tot} - \frac{1}{2}mE_{N_2}^{tot}}{n+m}, \qquad (1)$$

where the total energy for the YN$x_i$ point is $E_i^{tot}$, calculated using their respective cell. $E_{N_2}^{tot}$ is the energy of the free N$_2$ molecule and $E^{tot}_{Y\_hex}$ is the energy of the ground state of metallic yttrium (hexagonal). This definition of $E^f$ corresponds to the heat of formation per mol of the structure under consideration. The results are plotted in Fig. 3; the lower the energy, the more favorable the structure. The difference in heat of formation for yttrium (Y$_8$N$_0$) in hexagonal and cubic phases is barely visible in this *y*-scale, but it is 3 kJ mol$^{-1}$ lower for the hexagonal phase. For the next point (Y$_8$N$_1$) our calculations give -36.6 kJ mol$^{-1}$ for the cubic phase and -36.9 kJ mol$^{-1}$ for the hexagonal phase. Still, the hexagonal arrangement is energetically favored. However for the next data point (Y$_8$N$_2$), our calculations provide energies of -63.5 and -54.8 kJ mol$^{-1}$ for the cubic and hexagonal phases. There is a clear margin favorable for the formation of cubic yttrium nitrides, starting from $x \geq 0.25$. However for $x \leq 0.25$ the two phases may coexist since there is a marginal energy difference between the two Y-stackings.

The density of states (DOS) is an important information to understand the bonding and electronic properties of new compounds. We have calculated the DOS at different stoichiometries and symmetries. We can see in Figs. 4 and 5, from bottom to top, that there is gradual transformation of electronic structure as nitrogen is introduced in the yttrium matrix. In fact, the DOS for Y$_8$N$_1$ is comparable to the DOS of Y$_8$N$_0$. The difference between them is a narrow peak around 3 eV due to N-2*p* states and a small peak at 12.8 eV due N-2*s*. This fact agrees with our initial supposition that the structure of low-nitrogen yttrium nitrides can be approximate by the structure of metallic yttrium. The next observation is the formation of a narrow subband in the deep portion of the valence band (around -13 eV). For Y$_8$N$_1$ this band is barely present. As more nitrogen atoms are introduced in the Y matrix this band increases its importance. The partial DOS (not show here) reveal that this band is mainly due to N-2*s* states with small hybridization with Y-*s*, -*p* and –*d* states. There is an interesting energy shift for this subband. For low nitrogen concentration (Y$_8$N$_1$) this band is centered at -12.8 eV, so in cubic as in hexagonal phases. The band moves to -13.1 eV for Y$_4$N$_4$ and then moves back to -12.7 eV for stoichiometric YN. We interpret this shift in terms of charge redistribution between N and Y, providing an indication of the degree of ionicity of Y-N bonding. However, the absolute value of chemical shift in the N-2*s* position should be interpreted carefully because the reference point, i.e. Fermi level ($E_f$), also shifts with the nitrogen content.

The effect of nitrogen in the upper portion of the valence band (-5 to 10 eV) is evident in the Figs. 4 and 5. For $Y_8N_1$ the greater part of N $2p$ states goes into the bottom of the band (from -2.5 to -4 eV), while a slight portion goes to top (~ 5 eV). The contrast between Y and $Y_8N_1$ indicates that nitrogen acts as a scatter center: i.e. its main effect at low stoichiometries is to disperse the energy states of the metallic lattice. This tendency is to keep through the series. That is, the main portion of N $2p$ states goes the lower region of the valence band, while a small portion goes to the upper part; blue line in Figs. 4 and 5. At same time, the Y $4d$ and $5s$ states hybridize with the N $2p$ states. The nitrogen incorporation produces a splitting of Y $4d$ states in two regions where the main portion goes to the upper part of conduction band; red line in Figs. 4 and 5. As a result, the DOS in the valence region is characterized by two main crests and a valley between them. The depleted zone happens in the vicinity of $E_f$. For nonstoichiometric nitrides the DOS at $E_f$ is low, although nonzero. At $x = 1$ is the only point in our graph that the DOS is zero at $E_f$, but it happen only for cubic YN. Our calculations show a fundamental indirect band gap of ~0.3 eV, whilst screened-exchange (SX) local-density approximation (LDA) approach result in a gap of ~0.85 eV [8-9]. Although for hexagonal YN the present calculation exhibit a metallic character even at 1:1 stoichiometry.

**Discussion**

In this work we are assuming that the *hcp* and *fcc* structures of Y are undistorted; a presumption not necessarily valid in real compounds, where there are occupied and vacant interstices. Additionally, we have explored just two possible stacking for the Y-atoms, AB and ABC. However, there is a lot of possibilities of close packed stacking, and we cannot exclude the probable existence of other stable or metastable structures, as it is covered in Ref. 6. However, having investigated the atomic and electronic structure of the hexagonal and cubic Y-N in several stoichiometries, the result shows that interstitial N-atom reinforces the bonding in every one of the considered cases. Therefore, we think that this development is valid for all the possible stacking of Y. The reduced cell volume for the nitrides must entail higher Y-Y interactions since their distances decrease as $x$ increase. The decrease in cell volume is a peculiarity of group IIIB TMN [12, 21]. Intriguingly, the volume decrease in hexagonal phases is higher than in the cubic phases, whilst at same time, the corresponding bulk modulus is lower. Lower bulk moduli for same stoichiometries are a sign of weaker chemical bonds, and might be a consequence of N-N repulsive interactions, nearer in octahedral sites of *hcp* structures. Despite that, the tendencies in calculations are clear for both arrangements: decrease in volume and energy and increase in bulk modulus as increasing *x*. Since all the subsequent structures have a minimum against isotropic deformation and their heat of formation are lower than metallic yttrium, they should be considered metastable points and physical probable phases. To our knowledge, no hexagonal phases of YN have been reported to this date. A possible explanation of the lack of information on *hcp*-YN formation is that experimentally is easier to produce the *fcc* phase. However, for low-nitrogen contents the hexagonal phases are favored by a tiny margin. Thereafter, the two phases may coexist in the $x \leq 0.25$ region.

Another interesting point is the adaptation of the electronic band structure as nitrogen is incorporated in the Y-matrix. There is a strong hybridization between the N and Y states. However the N states goes preferably to the bottom of valence band, whilst a substantial number of states contributed by Y set in the top of valence band, causing a full splitting of the DOS at YN. Observe the DOS at $E_f$ in Figs. 4 and 5. To some extent, we are regulating the electronic population in the vicinity of $E_f$ with the incorporation of interstitial N. This must have important consequence in the conductivity (and other properties) of Y.

**Conclusions**

In summary, in this work we are showing a method to study the 'poisoning' of a metallic lattice by interstitial atoms. The nitrogen poisoning can be considered as a basic step in nitride formation. In particular, we have presented a systematic investigation of the electronic structure of cubic and hexagonal yttrium nitrides as a function of nitrogen occupation of octahedral interstices. Even though we are not sure that the suggested structures will be formed as the Y-N solid solution is made. We believe that much of their character will be similar to the one proposed here. We have covered the two important points for interstitial solid solutions. That is, one sequence, where the interstitial atom occupies shared faces, and another sequence, where the interstitial atom occupies shared edge octahedral sites. The calculation show that, for very low nitrogen incorporations the two phases may coexist. However, the phase transition to cubic symmetry is expected for $x \sim 0.25$. The cubic phase in 1:1 stoichiometry is favored by a 30 kJ mol$^{-1}$ margin.

Our calculations reveal that the DOS undergo a gradual modification as nitrogen is incorporated in the yttrium lattice. The effect of nitrogen is to split the valence band in two main peaks. The minimum is near $E_f$. For $x < 1$ the material behave as metal. At $x = 1$ the DOS is zero at $E_f$ for the cubic phase, confirming in this way the semiconductor character of YN. These results are important since the semiconducting property of *fcc*-YN could be utilized constructively in electronic devices, and maybe, even as active layer in diodes and transistors.

**Acknowledgements**

This work was supported by Supercomputer Center DGSCA-UNAM, DGAPA project IN120306 and CONACYT Grant U50203-F. We thank Juan Peralta, Margot Sainz, Carlos González, Eloisa Aparicio, and Jorge Palomares for technical support.

Table I. Transformation to lower the symmetry of cells in basic steps.

| Cubic cell transformation | | | Hexagonal cell transformation | | |
|---|---|---|---|---|---|
| 1 | 2 | 3 | 1 | 2 | 3 |
| Base group: *Fm3m* | Doubled Cell group: *Fm3m* | Reduced cell group: *R-3m* | Base group: *P63/mmc* | Double cell: *P63/mmc* | Reduced cell group: *P-3m1* |
| Axes: $x = y = z = a$; $\alpha=\beta=\gamma= 90°$ | Axes: $x = y = z = 2a$; $\alpha=\beta=\gamma=90°$ | Axes: $x' = \frac{1}{2} y + \frac{1}{2} z$; $y' = \frac{1}{2} x + \frac{1}{2} z$; $z' = \frac{1}{2} x + \frac{1}{2} y$; $\alpha=\beta=\gamma=60°$ | Axes: $x = y = a, z = c$; $\alpha=\beta= 90°$, $\gamma=120°$ | Axes: $x = y = 2a, z = c$; $\alpha=\beta= 90°$, $\gamma=120°$ | Axes: $x' = x$; $y' = y$; $z' = z$; $\alpha=\beta=90°$, $\gamma=120°$ |
| Wyckoff Y sites: 4*a* | Wyckoff Y sites: 24*e*, 8*c* | Wyckoff Y sites: 6*h*, 2*c* | Wyckoff Y sites: 2*c* | Wyckoff Y sites: 2*d*, 6*h* | Wyckoff Y sites: 2*d*, 6*i* |
| Wyckoff N sites: 4*b* | Wyckoff N sites: 4*a*, 4*b*, 24*d* | Wyckoff N sites: 1*a*, 1*b*, 3*d*, 3*e* | Wyckoff N sites: 2*a* | Wyckoff N sites: 2*a*, 6*g* | Wyckoff N sites: 1*a*, 1*b*, 3*e*, 3*f* |

Table II. N-filling of Wyckoff positions in the two groups to give the different stoichiometries.

| Formula in the unit cell | R-3m | | P-3m1 | |
|---|---|---|---|---|
| | Y Wyckoff positions | N Wyckoff positions | Y Wyckoff positions | N Wyckoff positions |
| $Y_8N_0$ | 6h, 2c | - | 2d, 6i | - |
| $Y_8N_1$ | | 1a | | 1a |
| $Y_8N_2$ | | 1a, 1b | | 1a, 1b |
| $Y_8N_3$ | | 3d | | 3e |
| $Y_8N_4$ | | 1b, 3d | | 1b, 3e |
| $Y_8N_5$ | | 1a, 1b, 3d | | 1a, 1b, 3e |
| $Y_8N_6$ | | 3d, 3e | | 3e, 3f |
| $Y_8N_7$ | | 1b, 3d, 3e | | 1b, 3e, 3f |
| $Y_8N_8$ | | 1a, 1b, 3d, 3e | | 1a, 1b, 3e, 3f |

Table III. - Cohesive energy and structural parameters of the calculated cells.

| Formula in the unit cell | $x_i$ | Cubic cell | | Hexagonal cell | |
|---|---|---|---|---|---|
| | | Cohesive Energy (eV) | $a_0^*$ (Å) | Cohesive Energy (eV) | $a_0^*, c_o^*$ (Å) |
| $Y_8N_0$ | 0.0 | -32.704 | 5.05437 | -32.931 | 3.63722, (3.6474)[†] 5.71811 (5.7306) |
| $Y_8N_1$ | 0.125 | -41.245 | 5.02935 | -41.276 | 3.60864, 5.67318 |
| $Y_8N_2$ | 0.25 | -49.312 | 4.98071 | -48.412 | 3.58816, 5.64099 |
| $Y_8N_3$ | 0.375 | -57.726 | 4.97468 | -57.315 | 3.57029, 5.61288 |
| $Y_8N_4$ | 0.50 | -66.724 | 4.95085 | -65.884 | 3.55722, 5.59234 |
| $Y_8N_5$ | 0.625 | -75.683 | 4.93055 | -74.009 | 3.53396, 5.55578 |
| $Y_8N_6$ | 0.75 | -84.698 | 4.91091 | -81.174 | 3.52531, 5.54217 |
| $Y_8N_7$ | 0.875 | -93.964 | 4.90968 | -90.250 | 3.51463, 5.52539 |
| $Y_8N_8$ | 1.0 | -102.906 | 4.90921 (4.877)[‡] | -98.342 | 3.51396, 5.52433 |

---

[*] Equivalent cell parameters to the reduced cubic and hexagonal phases.

[†] Experimental value, reference [18].

[‡] Experimental value, reference [19].

**Figures**

**(a)**

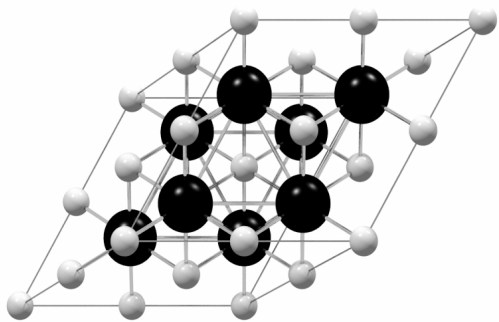

**(b)**

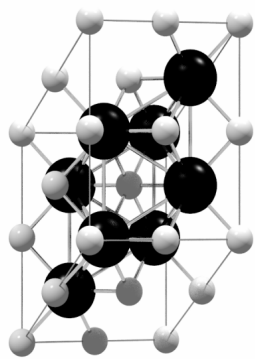

Fig. 1. Representation of the cubic (a) and hexagonal (b) phases of yttrium nitrides in *R-3m* and *P-3m1* cells, respectively. The metal atoms are the dark spheres.

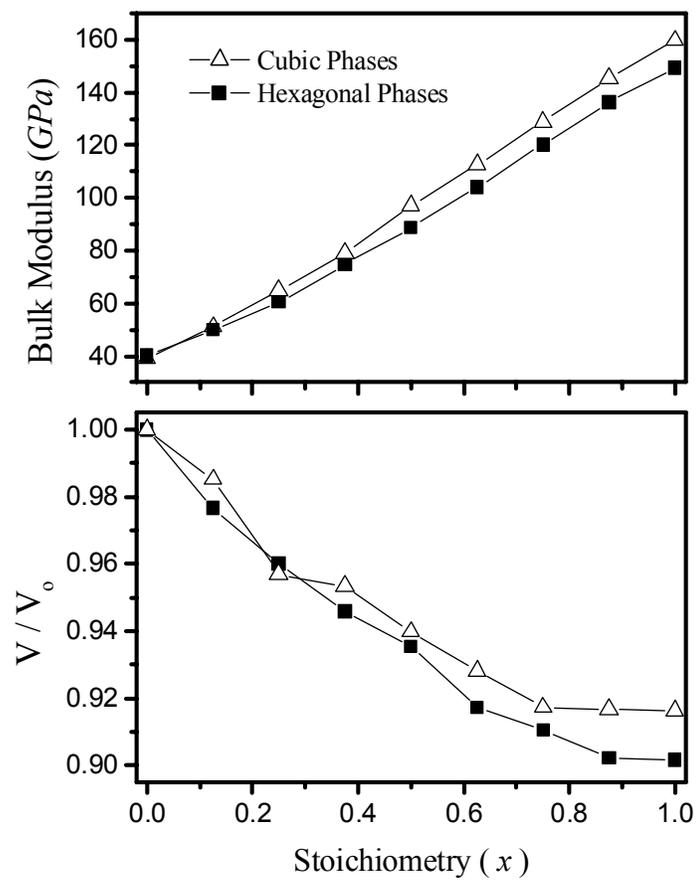

Fig. 2. Plot of the Bulk modulus and relative volume of Yttrium Nitrides in cubic (open triangles) and hexagonal (filled squares) symmetries.

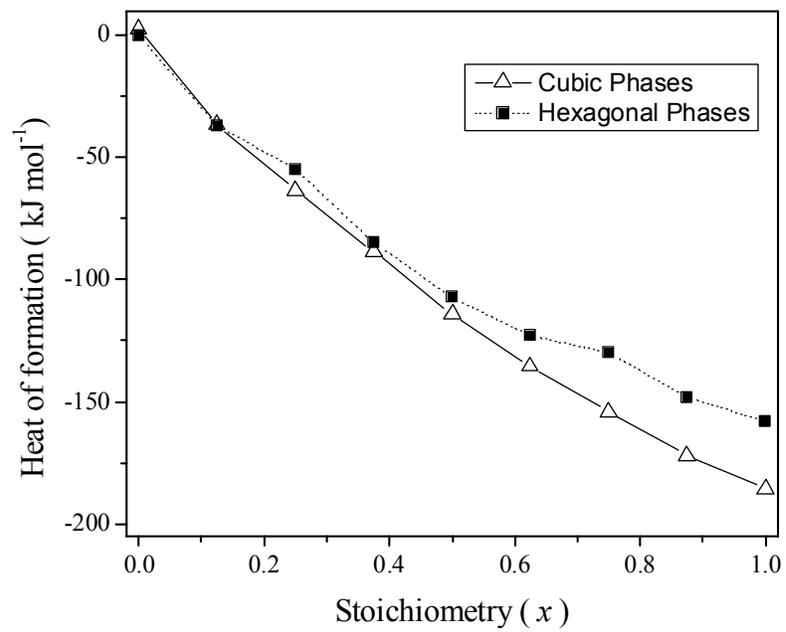

Fig. 3. Heat of formation of the cubic and hexagonal yttrium nitrides.

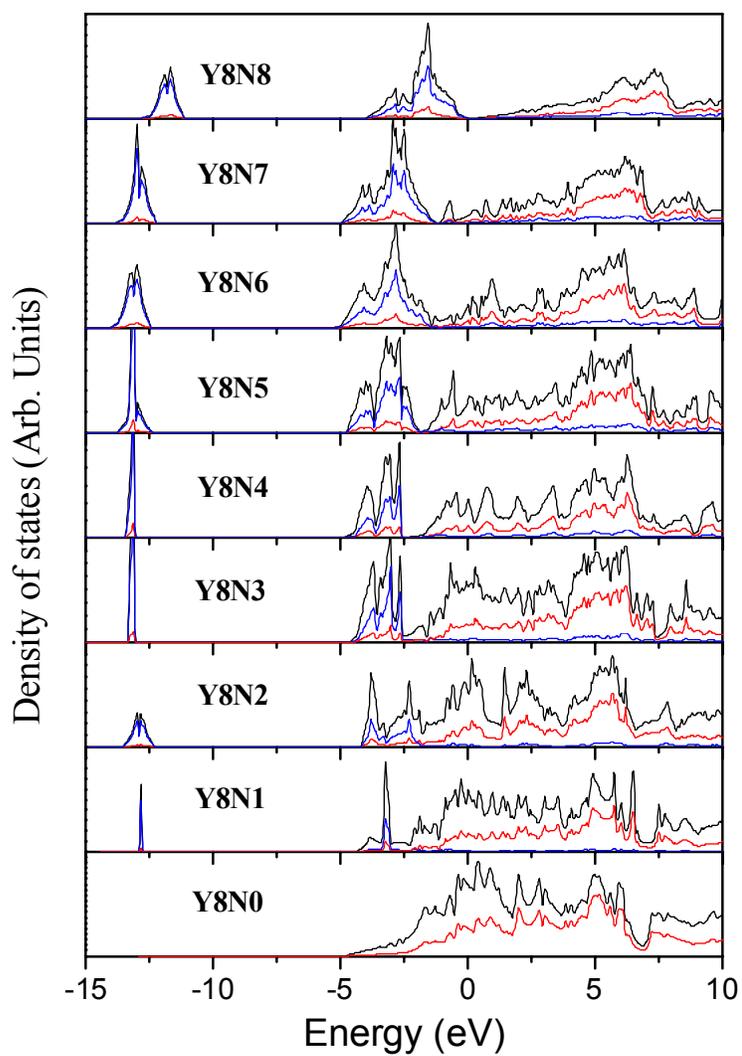

Fig. 4. Density of states of the calculated Yttrium nitride family in cubic symmetry. The Red line is the contribution of the Y atoms, while the blue line is the contribution of the N atoms.

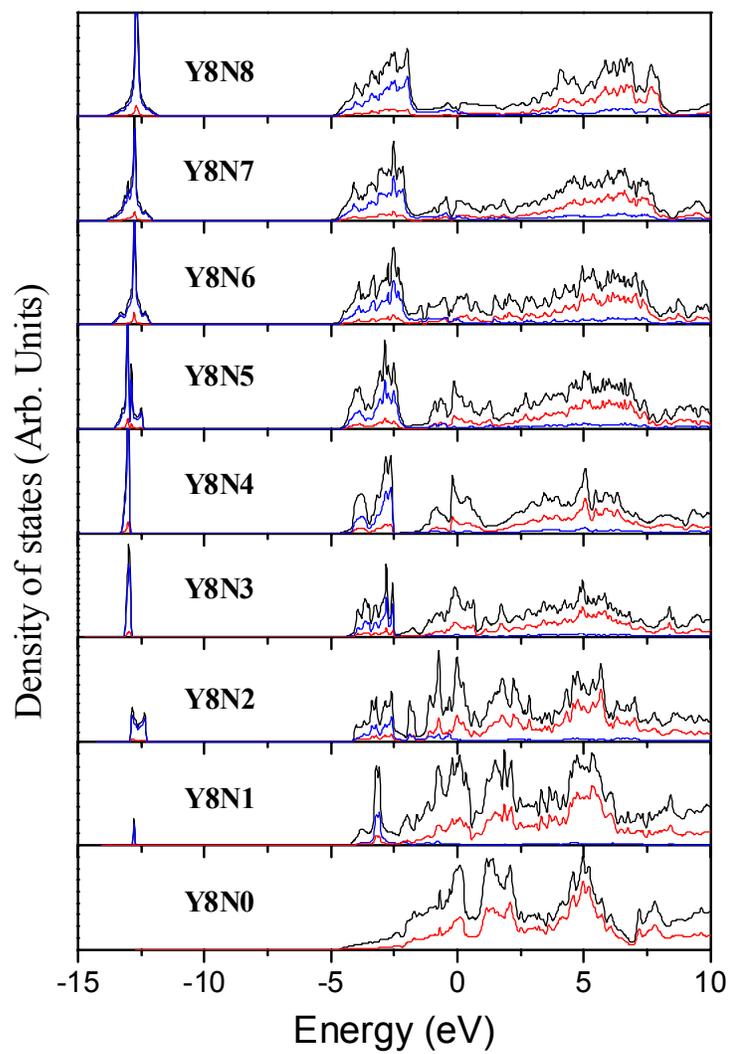

Fig. 5. Density of states of the calculated Yttrium nitride family in hexagonal symmetry. The Red line is the contribution of the Y atoms, while the blue line is the contribution of the N atoms.